\documentclass[aps,twocolumn,epsfig,eqsecnum,showpacs]{revtex4}
\usepackage{graphicx}

\def\cE{{\cal E}}
\def\cH{{\cal H}}
\def\cT{{\cal T}}
\def\cS{{\cal S}}

\def\T{{\rm Tr}}
\def\r{{\rangle}}
\def\l{{\langle}}

\begin{document}

\title{Approximate programmable quantum processors}
\author{Mark Hillery$^{1,2}$,M\'{a}rio Ziman$^{2,3}$, and Vladim\'{\i}r
Bu\v{z}ek$^{2,4}$}
\address{
$^{1}$~Department of
Physics, Hunter College of CUNY, 695 Park Avenue,  New York, NY 10021 USA\\
$^{2}$~Research Center for Quantum Information,  Slovak
Academy of Sciences, 845 11 Bratislava, Slovakia\\
$^{3}$~{\em Quniverse}, L{\'\i}\v{s}\v{c}ie \'{u}dolie 116,
841 04 Bratislava, Slovakia\\
$^{4}$~Abteilung f\"{u}r Quantenphysik, Universit\"{a}t Ulm, 89069 Ulm, Germany
}

\begin{abstract}
A quantum processor is a programmable quantum circuit in which both the data and the program,
which specifies the operation that is carried out on the data, are quantum states.  We study the
situation in which we want to use such a processor to approximate a set of unitary operators to
a specified level of precision.  We measure how well an operation is performed by the process
fidelity between the desired operation and the operation produced by the processor.  We show
how to find the program for a given processor that produces the best approximation of a particular
unitary operation.  We also place bounds on the dimension of the program space that is
necessary to  approximate a set of unitary operators to a specified level of precision.
\end{abstract}

\maketitle

%\begin{multicols}{2}
\section{Introduction}
Quantum circuits are typically designed to perform one function, for example, teleportation or
cloning.  It is useful to have circuits that are more flexible and can perform a variety of
functions.  The operation that the circuit performs can be determined either by classically setting the
values of some parameters, for example the rotation angle in a one-qubit rotation gate, or
it can be determined quantum mechanically, where a quantum system serves as a program to
tell the circuit what to do.  The second method has the advantage that the program could be a
result of a previous stage of a quantum computation, which would allow one stage of a
computation to control a subsequent one.

A programmable quantum circuit (quantum processor) has two inputs, the data register and
the program register.  The
data register is in the state on which we want to perform an operation, and the program state
specifies the operation.  An example is the C-NOT gate in which the control qubit is the program
and the target qubit is the data.  If the control qubit is in the state $|0\rangle$, nothing is done
to the data state, and if it is in the state $|1\rangle$, the operation $\sigma_{x}$ (bit flip) is
applied to the data state.  The superposition state $\alpha |0\rangle + \beta |1\rangle$ causes
the completely positive quantum map
\begin{equation}
T(\rho )= |\alpha |^{2} \rho + |\beta |^{2} \sigma_{x}\rho\sigma_{x} ,
\end{equation}
to be applied to the data state, $\rho =|\psi\rangle\langle\psi |$.

Suppose that we have a set of $N$ unitary operators that we want to be able to implement on the
data qubit with a programmable quantum circuit.  Nielsen and Chuang showed that this requires
a program space of at least $N$ dimensions \cite{nielsen1}.  This follows from the fact that  the
program states corresponding to any two of the unitary operators must be orthogonal.  If one
wants to be able to realize a large (or infinite) number of unitary operations with a program
space of fixed dimension, one has two possible options:  One option is to make the processor
probabilistic, that is a measurement is performed at the program output, and if the correct result
is obtained, the desired operation has been performed on the data \cite{nielsen1}-
\cite{buzek2}.  The probability of obtaining the proper measurement outcome will, in general, be
less than one so that the processor succeeds with only a certain probability.  The second option
is to make the processor an approximate one.  That is, each of the operations is not performed
exactly, but only up to some level of approximation.  It is this type of processors that we wish
to discuss here.

Approximate processors have been discussed by Vlasov \cite{vlasov}
and by Vidal and Cirac \cite{vidal}.  Vlasov considered a
classically programmable processor, while Vidal and Cirac
considered one whose programs are arbitrary quantum states.  They
made some rough estimates of the resources required  for a
processor to be able to program a set of unitary operators  to a
specified level of precision \cite{vidal}.
In Refs.\cite{dusek}-\cite{dariano}
approximate programmable quantum measurement devices
have been studied.  These devices  realize
certain classes of POVM's up to some level of approximation, and
which POVM they perform is determined by a program state.  It was
shown by D'Ariano and Perinotti that
for programmable measurement devices
the number of dimensions of the program space is a polynomial function of the reciprocal of
the desired accuracy \cite{dariano}.

Physical limitations in real systems lead to
an additional reason to study approximate processors. That is,
ideal devices in theory become approximate ones in practice.  For
example, if one wants to perform a rotation on a qubit that is a two-level atom, one applies
a classical field to the atom.  Real fields, however, consist of photons, and this and energy
constraints on the field place limits on the accuracy of the rotation that can be achieved
\cite{geabanacloche1}-\cite{geabanacloche2}.  Conservation laws can also place limits on
the accuracy of quantum operations \cite{ozawa}.

Other types of processors have been explored.  In particular, processors that evaluate the
expectation value of an arbitrary operator have been proposed \cite{ekert,paz}.  In these
processors, the data is the state in which the expectation value is to be evaluated, and the
program specifies the operator.

In this paper we shall discuss processors that approximate sets of unitary operators.  We shall
show, for a given processor, how to select an optimal program vector to approximate a particular
unitary operator.  In addition we shall give a lower bound on the number of dimensions the
program space must have to approximate a set of unitary operators to a given level of accuracy.
In the last part of the paper we will address the question
of optimal programmability, i.e. which processor is the best in
approximating all channels.

\section{Optimal program states}

We now consider a processor that acts on the Hilbert space
${\cal H}={\cal H}_{d}
\otimes {\cal H}_{p}$, where ${\cal H}_{d}$ is the data Hilbert space and
${\cal H}_{p}$ is the program Hilbert space.  Let us denote the dimension of
${\cal H}_{d}$ by $D$ and that of ${\cal H}_{p}$
by $N$. The processor itself is represented by a unitary operator $G$,
which acts on ${\cal H}$.  The action of the processor on the input
state $|\psi\rangle_{d} |\Xi\rangle_{p}$ is given by \cite{buzek3}
\begin{equation}
G(|\psi\rangle_{d}\otimes |\Xi\rangle ) =\sum_{j=1}^{N}A_{j}(\Xi )|\psi\rangle_{d}|j\rangle_{p} ,
\end{equation}
where $\{ |j\rangle_{p}| j=1,\ldots N\}$ is an orthonormal basis of
${\cal H}_{p}$.  The operators $A_{j}(\Xi )$ are expressed in terms
of the operators $A_{jk}$, where $G$ is expressed as
\begin{equation}
G=\sum_{j,k=1}^{N} A_{jk}\otimes |j\rangle_{p}\,_{p}\langle k|  .
\end{equation}
These operators obey the relations
\begin{eqnarray}
\label{Ajknorm}
\sum_{j,k=1}^{N}A_{jk_{1}}^{\dagger}A_{jk_{2}}= I_{d}\delta_{k_{1}k_{2}} \; ;\nonumber \\
\sum_{k=1}^{N}A_{j_{1}k}^{\dagger}A_{j_{2}k}= I_{d}\delta_{j_{1}j_{2}}\;  ,
\end{eqnarray}
where $I_{d}$ is the identity operator on ${\cal H}_{d}$.  The operator $A_{j}(\Xi )$ is given by
\begin{equation}
\label{Aj}
A_{j}(\Xi )=\sum_{k=1}^{N}A_{jk} \,_{p}\langle k|\Xi \rangle_{p}  ,
\end{equation}
from which it follows that
\begin{equation}
\label{Ajnorm}
\sum_{j=1}^{N}A^{\dagger}_{j}(\Xi ) A_{j}(\Xi ) = I_{d}  .
\end{equation}

We now need to discuss how to measure how close our processor comes to achieving a particular
unitary operation.  We shall use, what has been called by Gilchrist, et al., the process fidelity
\cite{nielsen2}, which was originally proposed by Raginsky \cite{raginsky}.  It is defined as
follows.  Let $T_{1}$ and $T_{2}$ be two completely positive maps, which map operators on
the Hilbert space ${\cal K}$ onto operators on the same space.  We shall assume that the
dimension of ${\cal K}$ is finite and equal to $D$.  The Jamiolkowski isomorphism allows
us to associate a density matrix on ${\cal K} \otimes {\cal K}$ with each of these maps.
Define the maximally entangled state
\begin{equation}
|\Phi\rangle = \frac{1}{\sqrt{D}}\sum_{j=1}^{D} |j\rangle |j\rangle ,
\end{equation}
where $\{ |j\rangle |j=1,\ldots N\}$ is an orthonormal basis of ${\cal K}$.  For each map
$T_{j}$, define the density matrix $\rho_{j}$ to be
\begin{equation}
\rho_{j} = ({\cal I}\otimes T_{j})(|\Phi\rangle\langle\Phi |) ,
\end{equation}
for $j=1,2$, where ${\cal I}$ is the identity map.  The process fidelity is defined as
\begin{equation}
F_{proc}(T_{1},T_{2})=\left[ {\rm Tr}\sqrt{\sqrt{\rho_{1}}\rho_{2}\sqrt{\rho_{1}}} \right]^{2} .
\end{equation}
The process fidelity has a number of useful properties that are discussed in Refs.
\cite{raginsky} and \cite{nielsen2}, one of which is the fact that it is symmetric, i.e.\
$F_{proc}(T_{1},T_{2})=F_{proc}(T_{2},T_{1})$.

We are going to be interested in the case in which one of the maps is unitary.  In particular, let
us assume that $T_{1}(\rho )=U\rho U^{-1}$ for some unitary operator $U$.  In this case we have
that $\rho_{1}$ is a pure state so that $\rho_{1}^{1/2}=\rho_{1}$.  This gives us that
\begin{equation}
{\rm Tr}\sqrt{\sqrt{\rho_{1}}\rho_{2}\sqrt{\rho_{1}}} =\frac{1}{D} \left[
\sum_{j_{1},j_{2}=1}^{D}\langle j_{1}|U^{-1}T_{2}(|j_{1}\rangle\langle j_{2}|)U|j_{2}\rangle
\right]^{1/2} .
\end{equation}
If $T_{2}$ is the result of the action of a processor, we have for a density matrix $\rho_{d}$,
representing a data state, that
\begin{equation}
T_{2}(\rho )=\sum_{j=1}^{N}A_{j}(\Xi )\rho_{d}A_{j}(\Xi )^{\dagger} ,
\end{equation}
which gives us, finally, that (we denote the map $T_{1}$ by the operator $U$)
\begin{equation}
\label{u_fidelity}
F(U,T_{2})=\frac{1}{D^{2}}\sum_{j=1}^{N} \left| {\rm Tr}(U^{-1}A_{j}(\Xi )) \right|^{2} .
\end{equation}
Using the notation for the Hilbert-Schmidt scalar product
$(A|B)={\rm Tr}A^\dagger B$ this can be rewritten in the form
$F(U,T_2)=\frac{1}{D^2}\sum_j |(A_j(\Xi)|U)|^2$.

This fidelity can also be expressed in terms of the operators $A_{jk}$.  Defining the matrix
\begin{equation}
M_{k_{1}k_{2}}=\frac{1}{D^{2}}\sum_{j=1}^{N}{\rm Tr}(A_{jk_{1}}^{\dagger}U){\rm Tr}
(U^{-1}A_{jk_{2}}) ,
\end{equation}
we have, from Eq.\ (\ref{Aj}), that
\begin{equation}
\label{FM}
F(U,T_{2})=\sum_{k_{1},k_{2}=1}^{N} \,_{p}\langle\Xi |k_{1}\rangle_{p} M_{k_{1}k_{2}}
\,_{p}\langle k_{2}|\Xi \rangle_{p} .
\end{equation}

Now consider the following problem.  Suppose we are given a processor and we wish to find the
best program to approximate the unitary operator $U$, where by best we mean the program that
maximizes the process fidelity.  An examination of Eq.\ (\ref{FM}) shows that this can be
accomplished by finding the eigenvector of
$M=\sum_{k_1,k_2}M_{k_1k_2}|k_1\rangle\langle k_2|$ with the largest eigenvalue, and choosing the
program vector to be this eigenvector.  The corresponding fidelity will just be the largest
eigenvalue of $M$.

This procedure is particularly simple to carry out when the processor is, what was called in
Ref. \cite{buzek3}, a U processor.  This is a processor that is a controlled-U gate.  Each
basis vector $|k\rangle_{p}$ in ${\cal H}_{p}$ is associated with a unitary operator $U_{k}$
acting on ${\cal H}_{d}$.  That is, if the program state is $|k\rangle_{p}$, then the operator
$U_{k}$ is applied to the data state.  The operators $A_{jk}$ for this type of processor are
particularly simple, $A_{jk}=\delta_{jk}U_{k}$, which implies that the matrix $M$ is given by
\begin{equation}
M_{k_{1}k_{2}}=\frac{1}{D^{2}}\left| \rm{Tr}(U^{\dagger}U_{k_{1}})\right|^{2}
\delta_{k_{1}k_{2}} .
\end{equation}
Because in this case $M$ is diagonal, we simply find the diagonal element that is largest.
This is the largest eigenvalue of $M$ and the maximum value of the fidelity.  The value of $k$ corresponding to this diagonal element tells us which of the basis vectors $|k\rangle_{p}$ is  the
program that will achieve this fidelity.  This implies that to best approximate a unitary operator
$U$ by a U processor, we simply find which of the unitary operators that the processor can
perform perfectly has the largest Hilbert-Schmidt inner product with $U$ and perform that
operation.  Note that this prescription does not make use of superpositions of the basis states
in the processor.
%%%%%%%%%%%%%%%%%%%%%%%%%%%%%%%%%%%%%%%%%%
\section{An example}
%%%%%%%%%%%%%%%%%%%%%%%%%%%%%%%%%%%%%%%%%%
Before proceeding with the exploration of the general properties of approximate quantum
processors, it is useful to analyze the following example.  We shall consider a processor acting
on qubits with an $N$ dimensional program space spanned by the orthonormal basis
$\{ |k\rangle_{p}|k=0, \ldots N-1 \}$.  Define the shift operators $E_{+}$ and $E_{-}$, acting on
the program space as $E_{+}|k\rangle = |k+1\rangle$ and $E_{-}|k\rangle = |k-1\rangle$,
where the addition and subtraction are modulo $N$.  We also define the program states
\begin{equation}
|\theta\rangle = \frac{1}{\sqrt{N}}\sum_{k=0}^{N-1}e^{-ik\theta}|k\rangle .
\end{equation}
If $\theta = \theta_{m} =(2\pi m)/N$ then the state $|\theta_{m}\rangle$ becomes an eigenstate
of $E_{+}$ and $E_{-}$
\begin{equation}
E_{+}|\theta_{m}\rangle = e^{i\theta_{m}}|\theta_{m}\rangle\; ;    \hspace{1cm}
E_{-}|\theta_{m}\rangle = e^{-i\theta_{m}}|\theta_{m}\rangle  .
\end{equation}
For the qubit, whose Hilbert space is spanned by the two orthonormal vectors $|0\rangle_{d}$
and $|1\rangle_{d}$, define the operators $\sigma^{(+)}$ and $\sigma^{(-)}$, where
$\sigma^{(+)}|0\rangle_{d} = |1\rangle_{d}$, $\sigma^{(+)}|1\rangle_{d} =0$, and
$\sigma^{(-)}=(\sigma^{(+)})^{\dagger}$.  We shall consider a specific realization of the U processor defined by the operator
$G$ acting on ${\cal H}_{d}\otimes {\cal H}_{p}$
\begin{equation}
G=\exp \left[ i \left( \frac{\pi}{2} \right) (\sigma^{(+)}\otimes E_{-}+\sigma^{(-)}\otimes E_{+}) \right]  .
\end{equation}
The fact that $G$ is a U processor can be seen when we let $G$ to act on the state
$|\psi\rangle_{d} |\theta_{m}\rangle_{p}$. Here we obtain the result
\begin{eqnarray}
|\Omega_m\rangle&=&G(|\psi\rangle_{d} \otimes
|\theta_{m}\rangle_{p})\\
\nonumber
&=&\exp \left[ i \left( \frac{\pi}{2} \right) \right.
\left.  (e^{-i\theta_{m}}\sigma^{(+)}+e^{i\theta_{m}}\sigma^{(-)})\right]|\psi\rangle_{d}\otimes
|\theta_{m}\rangle_{p}  .
\end{eqnarray}
Defining
\begin{equation}
U(\theta )=\exp \left[ i \left( \frac{\pi}{2} \right) \right.
\left.  (e^{-i\theta}\sigma^{(+)} +e^{i\theta}\sigma^{(-)})
\right]    ,
\end{equation}
we see that we can perform $U(\theta )$ perfectly when $\theta = \theta_{m}$, for some $m$.
Suppose, however, we are interested in using this processor to approximately perform
$U(\theta)$, for $\theta$ not equal to any of the $\theta_{m}$.  We know what the optimal
strategy is from the previous section, find the operator $U(\theta_{m})$ which has the
greatest overlap (in the sense of the Hilbert-Schmidt inner product) with $U(\theta)$ and perform
that operation.  Here we are going to examine a strategy, which is simpler to implement, but not
optimal.  We shall simply use the state $|\theta\rangle_{p}$ as a program state.  We find that
this gives us a process fidelity of
\begin{equation}
F=\frac{1}{N^{2}}\sum_{m=0}^{N-1} \cos^{2}(\theta_{m}-\theta )
\frac{\sin^{2}[N(\theta_{m}-\theta )/2]}{\sin^{2}[(\theta_{m}-\theta )/2]}  .
\end{equation}
This sum is an oscillatory function of $\theta$ with a period $2\pi/N$. The minima of this function
are achieved for $\theta=\pi/N + 2\pi k/N$ when the process fidelity takes the minimal value $F_{min}=1-2/N$.

Let us see how this compares to using the optimal program states.  The process fidelity between
the operators $U(\theta_{1})$ and $U(\theta_{2})$ is given by
\begin{equation}
F(U(\theta_{1}),U(\theta_{2}))=\cos^{2}(\theta_{1}-\theta_{2}) .
\end{equation}
If we approximate $U(\theta )$ by $U(\theta_{m})$, where $m$ is chosen so that $U(\theta )$
and $U(\theta_{m})$ have the largest Hilbert-Schmidt inner product, then the fidelity is
bounded below by
\begin{equation}
F\geq \cos^{2}\left(\frac{\pi}{N}\right) \sim 1- \left(\frac{\pi}{N}\right)^{2} .
\end{equation}
Note that in this case the error is of order $1/N^{2}$, while in the previous case it was of order
$1/N$, so there is a cost to not using the best program states.

 What we then have is a an approximate processor that can be made very accurate by choosing
$N$ large enough. It achieves an accuracy of order $1/N$ in approximating $U(\theta )$ with the simple program state $|\theta\rangle_{d}$, which is not as good as the best accuracy,
$1/N^{2}$, but the approximation in none the less a good one for $N$ sufficiently large.
Thus, we see that a U processor, making use of a simple program,  can be quite
useful in approximating the action of a set of operators labeled by a continuous parameter.

\section{Bound on dimension of program space}
We would now like to find a bound on the resources required to achieve a given accuracy in
approximating a set of unitary operators by means of a fixed processor.  In particular, we
want to see how the dimension of the program space grows as the accuracy of the
approximation increases.

The Schwartz inequality $|(A|B)|\le \sqrt{(A|A)(B|B)}$ implies that
\begin{equation}
|{\rm Tr}(U^{\dagger}A_{j}(\Xi ))| \leq \sqrt{D} [{\rm Tr}(A_{j}^{\dagger}(\Xi )A_{j}(\Xi ))]^{1/2}  ,
\end{equation}
and, therefore, if the action of our processor with the program state $|\Xi\rangle_{p}$ is given by
the map $T$, we have that
\begin{eqnarray}
F(U,T) & = & \frac{1}{D^{2}}\sum_{j=1}^{N} |{\rm Tr}(U^{\dagger}A_{j}(\Xi ))|^{2} \nonumber \\
 & \leq & \frac{1}{D} \sum_{j=1}^{N}{\rm Tr}(A^{\dagger}_{j}(\Xi )A_{j}(\Xi ))  = 1.
\end{eqnarray}
In the last equality we used the normalization property of Kraus
operators (\ref{Ajnorm}), i.e. $\sum_j A_j^\dagger(\Xi) A_j(\Xi)=I$.

We begin by assuming that the fidelity is $1$ and seeing what this implies about the operators
$A_{j}(\Xi )$.  If $F(U,T)=1$, then, we see from above, that Schwartz
inequality has to be saturated. This means that the
operators $A_j(\Xi)$ and $U$ are colinear, i.e. $A_j(\Xi)=\beta_j U$, where
$\beta_{j}$ is a complex number. Furthermore, Eq.\ (\ref{Ajnorm}) implies
$\sum_{j=1}^{N}|\beta_{j}|^{2}=1$. Now suppose that we have two
different unitary operators that can be realized perfectly, $U_{1}$
by the program state $|\Xi_{1}\rangle_{p}$ and $U_{2}$ by the program state $|\Xi_{2}\rangle$.
Therefore, $A_{j}(\Xi_{1})=\beta_{1j}U_{1}$ and $A_{j}(\Xi_{2})=\beta_{2j}U_{2}$.  We then
have that
\begin{eqnarray}
\sum_{j=1}^{N}\beta_{1j}^{\ast}\beta_{2j}U_{1}^{-1}U_{2} & = & \sum_{j=1}^{N}A_{j}^{\dagger}
(\Xi_{1})A_{j}(\Xi_{2})   \nonumber \\
 & = & I_{d} \,_{p}\langle\Xi_{1}|\Xi_{2}\rangle_{p}  ,
\end{eqnarray}
where we have used Eqs. (\ref{Aj}) and (\ref{Ajknorm}).  If $U_{1} \neq U_{2}$, then this equation
implies that both $\,_{p}\langle \Xi_{1}|\Xi_{2}\rangle_{p}$
and $\sum_{j=1}^{N}\beta_{1j}^{\ast}\beta_{2j}$ are zero.  This result is simply a restatement of the
Nielsen-Chuang theorem: If two unitary operators are realized
perfectly by a processor, their
program vectors must be orthogonal.

Now let us suppose that the processor performs the operation $U$ with a fidelity greater than
or equal to $1-\epsilon$, i.e. $F(U,T)\ge 1-\epsilon$, where $T$ is
specified by Kraus operators $A_j(\Xi)$. Let us express these operators as
\begin{equation}
\label{decomp}
A_{j}(\Xi )=\beta_{j}U+B_{j}(\Xi ) ,
\end{equation}
where ${\rm Tr}(U^{\dagger}B_{j}(\Xi ))=0$.  This decomposition is
unique. The inequality $F(U,T)\ge 1-\epsilon$ implies the following
condition on coefficients $\beta_j=\frac{1}{D}(U|A_j(\Xi))$
\begin{equation}
1\geq F(U,T)=\frac{1}{D^2}\sum_{j=1}^N
|(U|A_j(\Xi))|^2=\sum_{j=1}^N|\beta_j|^2\ge 1-\epsilon \, .
\end{equation}
Tracing both sides of the normalization condition $\sum_j A_j(\Xi)^\dagger A_j(\Xi)=I$
we obtain the inequality $\sum_j \T [B_j(\Xi)^\dagger B_j(\Xi)]=\sum_j (B_j(\Xi)|B_j(\Xi))\le D\epsilon$.

Next consider the situation in which our processor can approximate two unitary operators,
$U_{1}$ and $U_{2}$,
each with a fidelity greater than or equal to $1-\epsilon$.  In particular, if $T_{1}$ is the map
produced by the program state $|\Xi_{1}\rangle_{p}$ and $T_{2}$ is the map produced by
the program state $|\Xi_{2}\rangle_{p}$,   then both $F(U_{1},T_{1})$ and $F(U_{2},T_{2})$
are greater than or equal to $1-\epsilon$.  We also have that
\begin{eqnarray}
A_{j}(\Xi_{1}) & = & \beta_{1j}U_{1}+B_{1j}(\Xi_{1})\; ; \nonumber \\
A_{j}(\Xi_{2}) & = & \beta_{2j}U_{2}+B_{2j}(\Xi_{2})\;  ,
\end{eqnarray}
where ${\rm Tr}(U_{1}^{\dagger}B_{1j}(\Xi_{1}))={\rm Tr}(U_{2}^{\dagger}B_{2j}(\Xi_{2}))=0$.  As
in the case when the unitary operators were performed perfectly, consider the quantity
\begin{eqnarray}
\label{diffU}
I_{d}\langle\Xi_{1}|\Xi_{2}\rangle & = & \sum_{j=1}^{N}A_{j}(\Xi_{1})^{\dagger}A_{j}(\Xi_{2})
 \\
\nonumber
 & = & \sum_{j=1}^{N}[\beta_{1j}^{\ast}U_{1}^{\dagger}+B_{1j}^{\dagger}(\Xi_{1})] [\beta_{2j}U_{2}
+B_{2j}(\Xi_{2})]  .
\end{eqnarray}
Let us evaluate the absolute value of traces of both sides
\begin{eqnarray}
D|\langle\Xi_1|\Xi_2\rangle|&=& |\sum_j (A_j(\Xi_1)|A_j(\Xi_2))|\\
\nonumber
&=& |\sum_j [\beta^*_{1j}\beta_{2j}(U_1|U_2)+\beta^*_{1j}(U_1|B_{2j}) \\
\nonumber & & \ \ \ \ \ \ \ \ \
+ \beta_{2j}(B_{1j}|U_2)+(B_{1j}|B_{2j})] |\\
&\le& |(U_{1}|U_{2}) | \sum_j \beta^*_{1j}\beta_{2j} | +2D\sqrt{\epsilon}+D\epsilon .
\nonumber
\end{eqnarray}
In the last line we used the formulas
\begin{equation}
\begin{array}{rcl}
\sum_j |(B_{1j}|B_{2j})|&\le& \sum_j\sqrt{(B_{1j}|B_{1j})(B_{2j}|B_{2j})}\\
&\le& \sqrt{\sum_j (B_{1j}|B_{1j}) \sum_j (B_{2j}|B_{2j})}\\
&\le& D\epsilon \; ,
\end{array}
\end{equation}
and
\begin{equation}
\begin{array}{rcl}
|\sum_j \beta^*_{1j}(U_1|B_{2j})|&\le & \sum_j
 |\beta_{1j}|\sqrt{(U_1|U_1)(B_{2j}|B_{2j})}\\
&\le& D\sqrt{\epsilon}\; .
\end{array}
\end{equation}
As a result we obtain the bound on the inner product between two program states
\begin{equation}
|\langle\Xi_1|\Xi_2\rangle| \le \frac{1}{D} |(U_{1}|U_{2})| | \sum_j
 \beta^*_{1j}\beta_{2j} | +2\sqrt{\epsilon}+\epsilon\, .
\end{equation}
Next we will estimate the first term. The idea is to use
Eq.(\ref{diffU}) and apply both sides to a special vector
$|\psi_\eta\rangle$ that maximizes the quantity
$1-|\langle\psi |U_{1}^{\dagger}U_{2}|\psi\rangle |^{2}$.
Let us denote this maximum by $\eta$, i.e.
\begin{equation}
\eta=\max_\psi \left[ 1-|\l\psi|U_1^\dagger U_2|\psi\rangle |^2 \right]\, .
\end{equation}
This quantity describes the distinguishability of two unitary transformations,
and a short calculation shows that $\eta \leq \| U_{1} - U_{2}\|^{2}$.
After applying both sides of Eq.\ (\ref{diffU}) to $|\psi_{\eta}\rangle$ we find
the components of the resulting vectors orthogonal to $|\psi_{\eta}\rangle$
by applying the projection operator
$P_\eta^\perp=I-|\psi_\eta\rangle\langle\psi_\eta|$ to both sides. The left side
vanishes and we obtain the equality
\begin{equation}
\begin{array}{rcl}
0&=&P^\perp_\eta(\sum_j A_{j}(\Xi_{1})^\dagger A_{j}(\Xi_{2}) )|\psi_\eta\rangle\\
&=&\sum_j\beta_{1j}^*\beta_{2j} P_{\eta}^{\perp}U_{1}^{\dagger}U_{2} |\psi_\eta\rangle
+|\omega\rangle\; ,
\end{array}
\end{equation}
where
\begin{eqnarray}
|\omega\rangle & = & P_{\eta}^{\perp} \sum_{j=1}^{N} \left( \beta_{1j}^{\ast} U_{1}^{\dagger}
B_{2j}(\Xi_{2}) + \beta_{2j} B_{1j}^{\dagger}(\Xi_{1}) U_{2}    \right.  \nonumber \\
& & \left. + B_{1j}^{\dagger}(\Xi_{1}) B_{2j}(\Xi_{2}) \right) |\psi_{\eta}\rangle  .
\end{eqnarray}
We now want to find a bound on $\|\omega \|$.
Using the facts that the operator norm is bounded by the Hilbert-Schmidt norm, we have that
\begin{eqnarray}
\| \sum_{j=1}^{N} B_{j}^{\dagger}(\Xi_{1})B_{j}(\Xi_{2}) \| & \leq &  \sum_{j=1}^{N}
 (B_{j}(\Xi_{1})|B_{j}(\Xi_{1}))^{1/2} \nonumber  \\
  & & (B_{j}(\Xi_{2})|B_{j}(\Xi_{2}))^{1/2}  \nonumber \\
& \leq & \epsilon D ,
\end{eqnarray}
and
\begin{eqnarray}
\| \sum_{j=1}^{N} \beta_{1j}^{\ast} U_{1}^{\dagger} B_{2j}(\Xi_{2}) \| & \leq & \sum_{j=1}^{N}
|\beta_{1j}| (B_{j}(\Xi_{2})| B_{j}(\Xi_{2}))^{1/2}  \nonumber \\
& \leq & \sqrt{\epsilon D} .
\end{eqnarray}
Applying these inequalities we have that
$||\omega || \leq \epsilon D + 2\sqrt{\epsilon} D$.  In addition, we find that
$|| P_{\eta}^{\perp}U_{1}^{\dagger}U_{2} \psi_{\eta} || =\sqrt{\eta}$.  Therefore, we can conclude
\begin{equation}
\left| \sum_{j=1}^{N}\beta_{1j}^{\ast}\beta_{2j}\right|  \leq \frac{\epsilon D + 2\sqrt{\epsilon D}}
{\sqrt{\eta}} .
\end{equation}
Defining
\begin{equation}
F=\min \left( 1,\frac{\epsilon D+2\sqrt{\epsilon D}}{\eta}\right) ,
\end{equation}
we have, finally, that
\begin{equation}
|\langle\Xi_{1}|\Xi_{2}\rangle | \leq \frac{F}{D}|(U_{1}|U_{2})|+2\sqrt{\epsilon}+\epsilon .
\end{equation}
Note that in the case that both operations are carried out without error, in which case $\epsilon
=0$, this inequality implies that the program vectors must be orthogonal, recovering the known
result.

Now suppose that we have $M$ unitary operators that we want implemented by a processor
so that the process fidelity for each of the operators is greater than or equal to $1-\epsilon$.  How
many dimensions must ${\cal H}_{p}$ have?  In order to answer this question, we first find
the values of $Y_{jk}=(F/D)|(U_{j}|U_{k})|$ corresponding to each pair of operators in
our set, and use these values to find the largest set of linearly independent vectors in the
set of program vectors.  Linear independence can be deduced from the following result:
If $\{ v_{k}|k=1,\ldots K\}$ are vectors of length $1$, and
$|\langle v_{k_{1}}|v_{k_{2}}\rangle | < 1/(K-1)$, then the vectors $\{ v_{k}|k=1,\ldots K\}$ are
linearly independent \cite{vidal,buzek3}.  Suppose that there is a subset of our operators, with
$M^{\prime}$ members, whose pairs have small values of $Y_{jk}$, and let the largest value
of $Y_{jk}$ for this subset be $Y_{max}$. Then we have for all of the program vectors
corresponding to this set, that
\begin{equation}
|\langle\Xi_{j}|\Xi_{k}\rangle | \leq Y_{max}+2\sqrt{\epsilon}+\epsilon =q(Y_{max},\epsilon ) .
\end{equation}
Let $K_{q}$ be the largest integer such that
$K_{q}<(1/q)+1$.  What the result we just quoted implies, is that any set of vectors whose
size is $K_{q}$ or less, will be linearly independent.  Therefore, if $M^{\prime} \leq K_{q}$, then
all of the program vectors will be linearly independent, and the dimension of ${\cal H}_{p}$
must be at least $M^{\prime}$.  If $M^{\prime}>K_{q}$, then the dimension of ${\cal H}_{p}$
must be at least $K_{q}$.  This, then, is the restriction our result imposes on the dimension
of the program space.

As an example, suppose we want to implement the operators $I$, $\sigma_{1}$, $\sigma_{2}$,
and $\sigma_{3}$ on qubits, where the operators $\sigma_{j}$, for $j=1,2,3$,  correspond to
the usual Pauli matrices.  For all pairs of these operators we find that $Y_{jk}=0$, and
\begin{equation}
q(0,\epsilon ) = 2\sqrt{2\epsilon} +  \epsilon .
\end{equation}
Our bounds then give us that for $\epsilon <0.02$ the program space must have four
dimensions, for $\epsilon <0.05$ it must have at least three dimensions, and for
$\epsilon <0.17$ it must have at least two dimensions.

%%%%%%%%%%%%%%%%%%%%%%%%%%%%%%%%%%%%%%%%%%%%%%%%%%%%%%%%%%%%%%%%%%%%%%
\section{One-parameter group: Two approaches}
%%%%%%%%%%%%%%%%%%%%%%%%%%%%%%%%%%%%%%%%%%%%%%%%%%%%%%%%%%%%%%%%%%%%%%
Programmable processors can be exploited to implement quantum maps
probabilistically. In this case a specific measurement on the program state
is performed and if an {\it a priori} defined result is obtained then we know
that a desired operation has been performed on the data. In other words the
specific measurement that is accompanied by a post-selection induces
the desired transformation of the data register.
As was discussed in \cite{vidal} a probabilistic
processor without measurement can be used as an approximate processor.
In this case the transformation can be expressed as
\begin{equation}
\cE_\xi[\varrho]=p_{success}\cT[\varrho]+p_{error}{\cal N}[\varrho]\; ,
\end{equation}
where $\cT$ is the channel we want to approximate, and $p_{success}$ and $p_{error}$ 
are independent of the input data state, $\rho$.  Due to the concavity
of the square root of the process fidelity we find that $p_{success}\le F(\cE_\xi,\cT)$, i.e.
the accuracy of the approximation is bounded from below
by the probability of success.

Here we want to compare the performance of
a probabilistic processor used as an approximate
one with a different type of approximate processor
in order to see which requires greater
resources.  Both will be used to implement operators in the same one-parameter group.  In
particular, consider the operations on qudits (with orthonormal basis $\{ |k\rangle |k=1,\ldots D\}$)
specified by
\begin{equation}
U(\theta ) = e^{i\theta}|1\rangle\langle 1| +X ,
\end{equation}
where $X=\sum_{k=2}^{N} |k\rangle\langle k|$, and $0\leq \theta <2\pi$.

Consider the processor described by the operators $A_{jk}$
for $1\leq j,k\leq N$, where
\begin{equation}
A_{jk}= \left\{ \begin{array} {c} \delta_{jk}X+\delta_{k,j+1}|1\rangle\langle 1| \hspace{5mm} j<N\; ; \\
\delta_{Nk}X+\delta_{k,1}|1\rangle\langle 1| \hspace{5mm} j=N \; ,\end{array}  \right.
\end{equation}
originally described in Ref.~\cite{buzek2}.  With the program state
\begin{equation}
|\Xi\rangle = \frac{1}{\sqrt{N}}\sum_{k=1}^{N}e^{i(k-1)\theta}|k\rangle ,
\end{equation}
we find that for $1\leq j \leq N-1$
\begin{equation}
A_{j}(\Xi )=\frac{1}{\sqrt{N}}e^{i(j-1)\theta}U(\theta ),
\end{equation}
and for $j=N$
\begin{equation}
A_{N}(\Xi )= \frac{1}{\sqrt{N}}(e^{i(N-1)\theta}X+|1\rangle\langle 1|) .
\end{equation}
What this means is that if after the action of the processor, the program state is measured in
the basis $\{ |1\rangle_{p}, \ldots |N\rangle_{p} \}$ and if the result $|j\rangle_{p}$ is obtained, 
where $j\neq N$, then the
operation $U(\theta )$ has been carried out on the data. However,  if the result $|N\rangle_{p}$ is obtained,
then the operation $U(\theta )$ has not been performed.  Because each of these outcomes is equally likely, the probability of obtaining the
desired result is $(N-1)/N$.  If instead of measuring the output of the program register we
discard it, i.e.\ trace over it, we can use this processor as an approximate one.  The
process fidelity in this case is given by
\begin{equation}
F=1 - \frac{2(D-1)}{ND^{2}}(1-\cos (N\theta ))  .
\end{equation}

Another processor that will approximate this one-parameter group can be constructed by
dividing the interval $[0,2\pi )$ into subintervals and approximating all of the operators
$U(\theta )$ for $\theta$ in a particular subinterval by a single operator.  In particular,
let $\Delta\theta =\pi /N$, and approximate $U(\theta )$ for $2j\Delta\theta \leq \theta \leq
2(j+1)\Delta\theta$ by $U_{j}=U((2j+1)\Delta\theta )$, where $j=0,1,\ldots N-1$.  We now
define a U processor by setting, for $j,k=0,1,\ldots N-1$
\begin{equation}
A_{jk}=\delta_{jk}U_{j} .
\end{equation}
In order to approximate $U(\theta )$ for $2j\Delta\theta \leq \theta \leq 2(j+1)\Delta\theta$,
we choose the program state $|\Xi\rangle_{p}=|j\rangle_{p}$.  For this processor we find
that
\begin{equation}
1-F \leq \frac{2(D-1)}{D^{2}}(1-\cos\Delta\theta ) \sim  \frac{2(D-1)}{D^{2}} \frac{\pi^{2}}{4N^{2}} .
\end{equation}
By comparing the two fidelities, we see that for a fixed value
of the program space dimension,
$N$, the second processor will provide a greater accuracy.

%%%%%%%%%%%%%%%%%%%%%%%%%%%%%%%%%%%%%%%%%%%%%%%%%%%%%%%%%%%%%%%%%%%%%%
%%%%%%%%%%%%%%%%%%%%%%%%%%%%%%%%%%%%%%%%%%%%%%%%%%%%%%%%%%%%%%%%%%%%%%
\section{Conclusion}
We have examined the approximation of a set of unitary operators by means of a programmable
quantum circuit, i.e. a quantum processor.  The programs themselves are quantum states.  We
have shown, for a fixed processor, how to find the program that induces the
best approximation of a particular unitary operator.  In addition, we have found bounds on the size of the program space that is
necessary to approximate a set of operators to a given precision.

Approximate processors can be characterized by their accuracy and by the resources they
require.  By the accuracy, or level of precision, we mean the quantity
$\epsilon_G=1-\min_{\cE\in\Gamma}\max_{\xi\in \cS (\cH_p)} F(\cE,\cE_\xi)$ \cite{accuracy}. 
Here $\Gamma$ is the set of maps we want to realize, $\cS (\cH_p)$ is the set of positive
operators on $\cH_p$ with trace one (note that we are allowing mixed program states here)
and $\cE_\xi[\varrho]=\T_p G\varrho\otimes\xi G^\dagger$.
The dimension of the program space, $N$, characterizes the resources required. 
We wish to know how these two parameters are related.  We have made some progress here
in exploring this relation for limited sets of maps.  The problem becomes more difficult if
one considers $\Gamma$ to be the set of all unitary maps and harder yet if it is the set of all completely positive trace-preserving maps. Once we have these definitions of 
precision and resources, we can consider two problems.
First, given a specific degree of precision $\varepsilon_G$  for some set of maps $\Gamma$,
how large must the program space be?
Second, for fixed resources, what is the optimal processor, i.e. for which $G$ is the
accuracy the best ($\varepsilon_G$ the least)?
In Ref.~\cite{perinotti} one case of this problem was solved by D'Ariano and Perinotti.
The data states were qubits,
and $\Gamma$ was the set of unitary operators acting on a single qubit. The program 
space was also a single quibit so that $N=2$.  They then showed that
the optimal accuracy is given by $\varepsilon_G=3/4$. This precision can be achieved
when $G$ is a swap gate \cite{hbz}, i.e.
$C_{\tt SWAP}(|\psi\r\otimes|\phi\r)=|\phi\r\otimes|\psi\r$ for all
states $|\psi\r,|\phi\r$. For a processor with both the data and program spaces having the 
same dimension $D$ and $G$ given by the $d$-dimensional version of the
swap gate, we find that $F(U,\cE_\xi )$, where $\cE_\xi$ is the map induced on the data by
the processor with program $\xi$, is independent of both the program and $U$, and is 
equal to $1/D^2$.  This implies that for this processor, the accuracy is given by 
$\varepsilon_G=1-\frac{1}{D^2}$.
We suspect that this is the optimal value if the size of the
program register equals to the size of the data register, i.e. $N=D$, but whether this suspicion 
is correct is beyond the scope of this paper and will be analyzed elsewhere \cite{hbz}.

There are many open issues remaining.
One possibility is to shift our focus, and rather than ask
what type of the processor can perform a given set of operations with
a particular level of precision, ask
instead if it is possible to characterize the operations
that a given processor can perform to a
specified accuracy.  Another issue is the following.  So far, we have assumed that we are
approximating a set of unitary operators with just a single
use of a processor.  What happens if we
can use the same processor more than once?  It turns out that
multiple usage of the processor
 can significantly improve the accuracy of the approximation.
In particular, when the  $U$ processor (which can perform
a set of unitary operators perfectly)
is used $n$ times, the one can perfectly perform not only the original set of
operators, but any product of these operators that is of length $n$ or less.

It would also be
useful to find specific processors, which are not U processors, that can approximate a wide
class of unitary operations.  As we have seen, superpositions of the basis program states are
not useful in optimally approximating a unitary operator with a U processor, but they very
well may be useful in doing so with other types of processors.

Probabilistic processors have shown themselves to be very flexible devices.  They can perform
large classes of operations while requiring only limited resources.  Their drawback is that these
operations are performed with a probability that is less than one.  It remains to be seen how
flexible deterministic processors are, but the results here place some constraints on what they
can accomplish. In this paper we have given an example of how a probabilitic processor
can be used as an approximate one.

\acknowledgements
This work
was supported in part by the European Union  projects QGATES, QAP, and
CONQUEST,  by the Slovak Academy of Sciences via the project CE-PI under the contract I/2/2005,
and by the project APVT-99-012304. VB thanks the Alexander von Humboldt Foundation for a
support.

%\end{multicols}


\begin{thebibliography}{99}
\bibitem{nielsen1} M.\ Nielsen and I.\ Chuang, Phys.\ Rev.\ Lett.\ {\bf 79}, 321 (1997).
\bibitem{vlasov}A.\ Yu.\ Vlasov, quant-ph/0103119.
\bibitem{vidal} G.\ Vidal and J.\ I.\ Cirac, quant-ph/0012067.
\bibitem{buzek1} M.\ Hillery, M.\ Ziman, and V.\ Bu\v{z}ek, Phys.\ Rev.\ A {\bf 65}, 022301 (2002).
\bibitem{buzek2} M.\ Hillery, M.\ Ziman, and V.\ Bu\v{z}ek, Phys.\ Rev.\ A {\bf 69}, 042311 (2004).
\bibitem{dusek} M.\ Du\v{s}ek and V.\ Bu\v{z}ek, Phys.\ Rev.\ A {\bf 66}, 022112 (2002).
\bibitem{fiurasek} J.\ Fiura\v{s}ek, M.\ Du\v{s}ek, and R.\ Filip, Phys.\ Rev.\ Lett.\ {\bf 89},
190401 (2002).
\bibitem{dariano} G.\ M.\ D' Ariano and P.\ Perinotti, quant-ph/410169.
\bibitem{geabanacloche1} J.\ Gea-Banacloche, Phys.\ Rev.\ A {bf 65}, 022308 (2002).
\bibitem{deutsch} A.\ Silberfarb and I.\ Deutsch, Phys.\ Rev.\ A {\bf 68}, 013817 (2003).
\bibitem{itano} W.\ Itano, Phys.\ Rev.\ A {\bf 68}, 046301 (2003).
\bibitem{geabanacloche2}  J.\ Gea-Banacloche, Phys.\ Rev.\ A {bf 68}, 046303 (2002).
\bibitem{ozawa} M.\ Ozawa, Phys.\ Rev.\ Lett. {\bf 89}, 057902 (2002).
\bibitem{ekert} A.\ K.\ Ekert, C.\ M.\ Alves, D.\ K.\ L.\ Oi, M.\ Horodecki, P.\ Horodecki, and
L.\ C.\ Kwek, Phys.\ Rev.\ Lett.\ {\bf 88}, 217901 (2002).
\bibitem{paz} J.\ P.\ Paz and A.\ Roncaglia, quant-ph/0306143.
\bibitem{buzek3} M.\ Hillery, M.\ Ziman, and V.\ Bu\v{z}ek, Phys.\
  Rev.\ A {\bf 66}, 042302 (2002).
\bibitem{ziman03} M.\ Ziman and V.\ Bu\v zek, Int. J.Quant. Inf. {\bf 1}, 523 (2003)
\bibitem{nielsen2} A.\ Gilchrist, N.\ K.\ Langford, and M.\ A.\ Nielsen, quant-ph/0408063.
\bibitem{raginsky} M.\ Raginsky, Phys.\ Lett.\ A {\bf 290}, 11 (2001).
\bibitem{accuracy}
The definition of the accuracy of the processor
$\varepsilon_G$ is not unique. For instance, one can consider the CB-norm
and use averages in the definition.
\bibitem{perinotti}
G.\ M.\ D'Ariano and P.\ Perinotti, quant-ph/0510033
\bibitem{hbz}
M.\ Hillery, V.\ Bu\v zek, and M.\ Ziman,
in preparation
\end{thebibliography}
\end{document}